\documentclass[referee,sn-mathphys-num]{sn-jnl}
\usepackage{graphicx}%
\usepackage{multirow}%
\usepackage{amsmath,amssymb,amsfonts}%
\usepackage{amsthm}%
\usepackage{mathrsfs}%
\usepackage[title]{appendix}%
\usepackage{xcolor}%
\usepackage{textcomp}%
\usepackage{manyfoot}%
\usepackage{booktabs}%
\usepackage{algorithm}%
\usepackage{algorithmicx}%
\usepackage{algpseudocode}%
\usepackage{listings}%
\usepackage{afterpage}
\usepackage{ulem}
\usepackage[export]{adjustbox}
\usepackage[english]{babel}

\begin{document}

\title{Integrated nano electro-optomechanical spiking neuron}

\author[1]{\fnm{Gregorio} \sur{Beltramo}}
\author[1]{\fnm{Róbert} \sur{Horváth}}
\author[1]{\fnm{Grégoire} \sur{Beaudoin}}
\author[1]{\fnm{Isabelle} \sur{Sagnes}}
\author[1]{\fnm{Sylvain} \sur{Barbay}}
\author*[1,2,3]{\fnm{Rémy} \sur{Braive}}\email{remy.braive@c2n.upsaclay.fr}

\affil[1]{\orgdiv{Centre de Nanosciences et Nanotechnologies}, \orgname{Université Paris Saclay}, \orgaddress{ \city{Palaiseau}, \postcode{91120}, \country{France}}}

\affil[2]{\orgname{Université Paris Cité}, \orgaddress{\city{Paris}, \postcode{75006}, \country{France}}}

\affil[3]{ \orgname{Institut Universitaire de France}, \orgaddress{\city{Paris}, \postcode{75231}, \country{France}}}

\abstract{

Neuromorphic computing offers a pathway toward energy-efficient processing of data, yet hardware platforms combining nanoscale integration and multimodal functionality remain scarce. Here we demonstrate a gallium-phosphide electro-optomechanical spiking neuron that integrates optical and electromechanical interfaces within a single nanostructure on a silicon photonic chip operating at telecommunication wavelengths (1550 nm) and exploiting a 3 gigahertz-frequency mechanical mode. Our device displays excitable dynamics, generating optical spikes at its output, as in the spiking activity of neurons and cardiac cells and defined by the calibrated all-or-none response to external perturbations. This dynamic is consistent with the saddle-node on invariant circle scenario and associated features are demonstrated including control of excitable threshold, temporal summation and refractory period. Our device compact footprint and its CMOS-compatible platform make it well suited for edge-computing applications requiring low latency and establish a foundation for versatile brain-inspired optomechanical computing and advanced on-chip optical pulse sources.

%the device displays excitable dynamics generating optical spikes at its output, as well as other brain-inspired properties such as temporal summation and refractory periods. This nano-optomechanical neuron enables fully integrated, spike-based processing of optical and mechanical signals and establishes foundations for energy-efficient on-chip generation of arbitrary optical spike sequences. Its compact footprint and CMOS-compatible silicon platform make it well suited for edge-computing applications requiring low latency and reduced bandwidth. This work establishes a versatile and novel nano-optomechanical platform for potentially efficient  neuromorphic and analogue computing.
}

%\section{Images}
\maketitle

%% Nature Nano 3000 mots (sans les methodes, 3000 mots aussi)
%% Up to 6 figures/tables
%% abstract 150 mots
%%% ~50 refs max

%\section{Introduction}
Among the numerous possibilities for building artificial neural networks hardware \cite{MarkovicNRP20}, those based on biologically inspired spiking neurons form the third generation of neural networks \cite{MAASS19971659}. Neuromorphic computers can perform computation through directed graphs that are
better suited for the collocation of computing units and memory much like the human
brain. Each neuron is an individual computing unit with its own dedicated memory
such that multiple pieces of information can be processed completely asynchronously
and in parallel \cite{brunner2025}. Within this framework, spiking or excitable neurons act as a building blocks for brain-inspired, neuromorphic computing architectures, offering new ways to represent information that may be more natural for specific classes of computation, such as graph algorithms \cite{chou_et_al:LIPIcs.ITCS.2019.26}, quadratic programming \cite{10.1162/neco_a_01113}, and other parallelizable computation algorithms \cite{Feldmann2021}. Encoding of information in the form of spike timing (temporal coding) or the spike rate (rate coding) has been investigated and subject to active research \cite{GAUTRAIS199857,StocklNMI21}. Excitable neurons are also attracting increasing interest because of their ability to compute a task using minimal resources, hence enabling potentially less energy consumption \cite{GoltzNMI21,RaoNMI22}. 

A hallmark of an excitable behaviour is the calibrated, all-or-none response to a perturbation. An excitable system possesses one stable state. When subject to a perturbation, the system remains in its quiet, stable state if the perturbation amplitude is smaller than a certain threshold.  If the perturbation amplitude exceeds this threshold (the excitable threshold), the system performs a characteristic excursion in phase space generally associated to the generation of a pulse and returns to its stable state.  Excitability has been demonstrated in many physical systems, including memristors \cite{Thomas_2013}, spintronic oscillators \cite{MarkovicPRB22}, Josephson junctions \cite{Mishra_2021}, microfluidics \cite{PedaciNP11} and lasers or optical systems \cite{PlazaEPL97,GiudiciPRE97,WunschePRL02,BarlandPRE03,GouldingPRL07,BeriPLA10,BarbayOL11}. One of the possible underlying physical mechanisms corresponds to a Saddle Node on Invariant Circle (SNIC) bifurcation \cite{Izhikevich_2006}. From a nonlinear dynamical point of view, the phase dynamics of a nonlinear oscillator can be generically represented by an invariant circle encircling the origin in phase space. In the unlocked regime, the oscillator phase rotates freely on the circle. However, in the injection-locked regime, the system's phase settles to a fixed stable point that appears on the circle together with an unstable fixed point, following a saddle-node bifurcation.
When perturbed beyond the minimum distance between these two fixed points,  the system follows a deterministic and repeatable trajectory in phase space to return to its stable state after completing a full $2\pi$ phase excursion. Usually associated to a spike in the amplitude response, this specific property can be found in injected single-mode oscillators, near the locked - unlocked transition and characterizes the excitable behaviour. 

Optoelectronic implementations of artificial spiking neural networks have been developed to combine the advantage of well-studied electronic nonlinearities alongside the fast, lossless transmission and zero-energy weighting provided by the photonic devices 
%Opto-electronic Oscillators (OEO) convert the aggregated synaptic optical inputs into electrical currents. The electrical circuit, in turn, can use any nonlinear circuit element to implement the spiking function and generate an optical output using external lasers 
\cite{Rohlev:91, RomeiraOE13, Chen_2023, Li_2024}. 
However, such a neuron architecture necessitates the use of electronic/optical conversions hence a potential loss of speed, increased complexity and thus increased energy consumption. Despite the integration of CMOS transistors with photodiodes and electro-optic modulators, the response speed of a single optoelectronic neuron is limited because of the analogue bandwidth of the electronics, unlike all-optical neurons.
%All-optical counterpats rely on optical nonlinearities found in the constitutive materials (intrinsic, opto-thermal) and on excitation arrangement (coherent injection, optical feedback, incoherent pumping) \cite{Duport:12, Shastri2021}. In semiconductors, they have been demonstrated in semiconductor optical amplifiers (SOAs) \cite{Garbin_2017}, vertical-cavity surface-emitting lasers (VCSELs) with injected signal \cite{Goulding_2007}, VCSEL with saturable absorber \cite{Selmi_2014}, passive micro-resonators \cite{Brunstein_2012, BERI2010739} to name a few.
%However, throughput increases in neural networks benefit from optical parallelism in wavelength and spatial domains.
More recently, Micro/Nano Electro-Mechanical Systems (MEMS/NEMS) have been considered for classical echo-state (non-spiking) reservoir computing \cite{Guo2024, Dion2021, Jin2022} and for image classification using the nonlinear transfer function of the mechanical response \cite{Dion_2018}. At last, a MHz thermo-optomechanical neuron-like behaviour has been reported in a silicon microcavity \cite{Yang_2020}.
Due to the coexistence on the same device of photonic and mechanical degrees of freedom, optomechanical resonators \cite{Aspelmeyer_2014} have the advantage of their all-optical counterpart and of naturally generating a radio-frequency (RF) signal like in Opto-electronic Oscillators (OEO). Besides quantum technologies developments \cite{Barzanjeh2022}, they present an extraordinary and yet scarcely explored platform for neuromorphic architectures incorporating spiking artificial neurons.

We demonstrate here the excitable behavior of a self-sustained, 3.1 GHz electro-optomechanical oscillator (E-OMO) made of Gallium Phosphide (GaP) and integrated on a Silicon-on-Insulator (SoI) photonic chip working at 1550 nm. Thanks to the piezoelectricity of GaP, injection locking of the electro-optomechanical oscillator to an external RF driving signal is achieved through on-chip integrated electrodes. External optical perturbations at the optical input port allow to reveal the excitable response. A statistical analysis is performed as a function of the perturbation strength as well as the RF drive detuning. Calibrated all-or-none responses are experimentally evidenced thus demonstrating the excitable behavior of the E-OMO. Beyond excitability, other neuromorphic features such as temporal summation and refractory periods are observed which pave the way to a novel integrated, analog neuromorphic computing platform.

\section{Electromechanical injection locking}

\begin{figure}[h]
    \centering
    \includegraphics[width=1\linewidth]{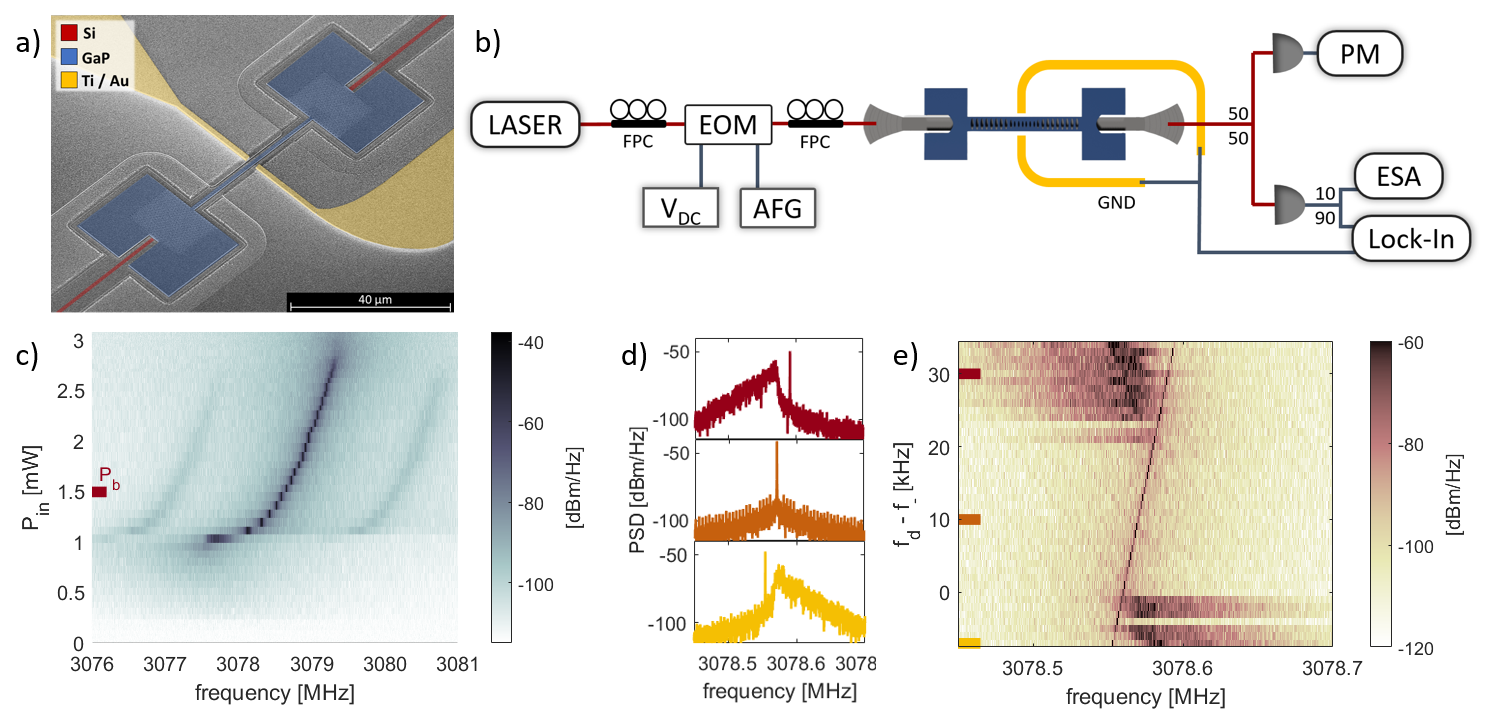}
    \caption{a) Colored SEM image of the electro-optomechanical oscillator showing the piezoelectric Gallium Phosphide photonic crystal nanobeam cavity (in blue) suspended on top of a silicon-on-insulator bus waveguide (in red) and with gold electrodes (in yellow) positioned on both sides of the PhC nanobeam cavity. 
    b) Experimental setup: tunable fibered IR laser (LASER), fibered polarization controller (FPC), $10$ GHz bandwidth electro-optic modulator (EOM), bias voltage supply (V$_{DC}$) and arbitrary function generator (AFG) constitute the excitation scheme. After the electro-optomechanical oscillator, the optical signal is split in two ports (50/50) and directed to an optical powermeter (PM) and a fast, 3.5 GHz bandwidth, photo-detector. The photodetected electrical signal is split 90\% to the lock-in amplifier and 10\% to the electric spectrum analyzer (ESA). The digital lock-in amplifier generates the RF driving at frequency $f_{d}$ at the gold electrodes. 
    c) Power spectral density map of the photodetected signal recorded at the ESA as a function of the input power $P_{in}$. The highlighted power level $P_{in}=P_{b}=1.51$ mW corresponds to the one used for excitability measurements.
    d) Power spectral densities  for $P_{in}=P_{b}$ and for three distinct frequencies offsets $f_{d}-f_{-}=-7, \ 10,\ 30$ kHz: before (bottom - yellow), inside (middle - orange) and after (top - red) the locking region, respectively. In c-d), $V_{RF} = 100$ mV. 
    e) Full power spectral density map of the photodetected signal as a function of the driving frequency offset for the same parameters as in c-d).  
    %The OMO frequency increases with the input power in a non-linear way, it rapidly increases from $P_{in}=0.6$ mW to $1.2$ mW after which it saturates around $3078.8$ MHz. The red dotted line corresponds to $P_{in}=1.05$ mW, the working point of the experiment. By locking the OMO to the driving frequency at $P_{in}=1.05$ mW and at $f_{drive} \simeq f_{locking}$ a short perturbation which increases the input power (vertical excursion) will push the OMO to unlock for the duration of this perturbation.
    }
    \label{fig:Fig1_SetUp_Lockin}
\end{figure}

We first discuss the electromechanical injection locking of our E-OMO. Our device consists of a 300 nm-thick gallium phosphide (GaP) and free-standing 1D optomechanical crystal (in blue in figure \ref{fig:Fig1_SetUp_Lockin}a)  heterogeneously integrated on top of a SoI circuitry (See Methods for more details). The control of the position of the holes along the 1D optomechanical  crystal structure allows to simultaneously confine photons around 1550 nm and phonons close to 3.1 GHz, giving rise to a large optomechanical coupling constant. The optomechanical  crystal is optically addressed through an underneath SoI optical waveguide (in red in figure \ref{fig:Fig1_SetUp_Lockin}a). 
A tunable laser, coupled to the SoI waveguide thanks to a grating coupler, is put on resonance with the optical mode of the 1D nanobeam at 1550.5 nm. Light is then coupled out trough another grating coupler to be analyzed. At low input optical power $P_{in}$ (measured before the input to the grating coupler), a fine scanning of the input wavelength gives access to a high resolution transmission spectrum as measured  with a power meter (PM in fig. \ref{fig:Fig1_SetUp_Lockin}b) and hence to the total optical quality factor ($Q_o$) of about $44\times10^3$. For $P_{in} = 80 \ \mu$W, and at $\lambda_o = 1550.5$ nm, output light is characterized by a fast photo-diode and analyzed with an electrical spectrum analyzer (ESA - Fig. \ref{fig:Fig1_SetUp_Lockin}b). The mechanical response of the optomechanical crystal is assessed under atmospheric conditions, allowing one to extract the mechanical resonant frequency $\Omega_m/ 2\pi = 3.078$ GHz, the mechanical quality factor $Q_m \approx 1550$ and the optomechanical coupling constant $g_0 / 2\pi = 600$ kHz.
By increasing input optical power $P_{in}$, self-sustained optomechanical oscillations \cite{Horvath_2023} appear above a given threshold and are evidenced by the sharp increase of the amplitude of the investigated mechanical mode (Fig. \ref{fig:Fig1_SetUp_Lockin}e). A non-linear frequency shift with optical power is also clearly visible because of the competition between optical spring effect and thermo-mechanical shift \cite{Jiang2016}. Nevertheless, the positive shift suggests that the the former is the main factor at play here. The two frequency sidebands visible in the mechanical RF spectrum are related to the mixing of the self-oscillating mode with a low frequency mechanical mode at $2$ MHz.  In the following, we set the injected optical power at $P_{in} = P_{b} \equiv 1.51$ mW. 

Electromechanical control of the mechanical mode \cite{Bekker:17, Horvath_2023} is obtained thanks to the piezoelectric properties of GaP, using 1 µm wide metallic electrodes (in yellow in figure \ref{fig:Fig1_SetUp_Lockin}a) located near the center of the nanobeam cavity.
A lock-in amplifier is used to inject a RF drive at a fixed peak-to-peak voltage $V_{RF} = 100$ mV. By sweeping its frequency $f_{d}$ upwards in the vicinity of the mechanical resonance $\Omega_m/2\pi$, different responses are visible on the ESA (Fig. \ref{fig:Fig1_SetUp_Lockin}d). For $f_d$ away from $\Omega_m/2\pi$, two peaks appears on the spectrum: the sharp one corresponds to the RF drive whereas the larger one is the mechanical mode distorted by the presence of the drive. Note that depending on the sign of the detuning $\Omega_d-\Omega_m$, the large spectral feature displays a distortion in opposite directions, due to the nonlinear frequency mixing between the self-induced optomechanical oscillation and the RF drive. By contrast, when the drive is closer to the natural frequency of the mechanical self-oscillator $\Omega_m/2\pi$, only a single peak is present (middle panel in figure \ref{fig:Fig1_SetUp_Lockin}d). 
This evidences injection locking of the mechanical mode with the external RF drive. By recording responses over a larger frequency range, the full electro-mechanical injection locking region can be observed with a width of 23 kHz (Fig. \ref{fig:Fig1_SetUp_Lockin}e) for the specific parameters chosen here ($V_{RF} = 100$ mV and $P_{b} = 1.51$ mW). The low and high frequency edges of the locking region are named $f_{-}$ and $f_{+}$ respectively in the following. 
%The width of the locking region is investigated as function of the optical input power $P_{in}$ (Fig. \ref{fig:Fig1_SetUp_Lockin}e). For the sake of simplicity, the response has been binarized between locked (black) and unlocked (salmon-colored) regimes. Here, two features are highlighted as $P_{in}$ increases. The locking region shifts towards higher frequency due to the well-known optical spring effect on the mechanical mode. As the optical power increases, self-sustained oscillations of the mechanical mode get larger so that the effective force applied is stronger. Thereby the locking region widens. In the following, $P_{in}$ is set at 0.56 mW highlighted by the red dashed line in figure \ref{fig:Fig1_SetUp_Lockin}e. Evolution of locking region as function of $V_{RF}$ is not shown here.{\color{red} Is it somewhere ? In the supplementary ? Otherwise remove.}

\section{Excitability of the Electro-optomechanical oscillator }

\begin{figure}
    \centering
    \includegraphics[width=1\linewidth]{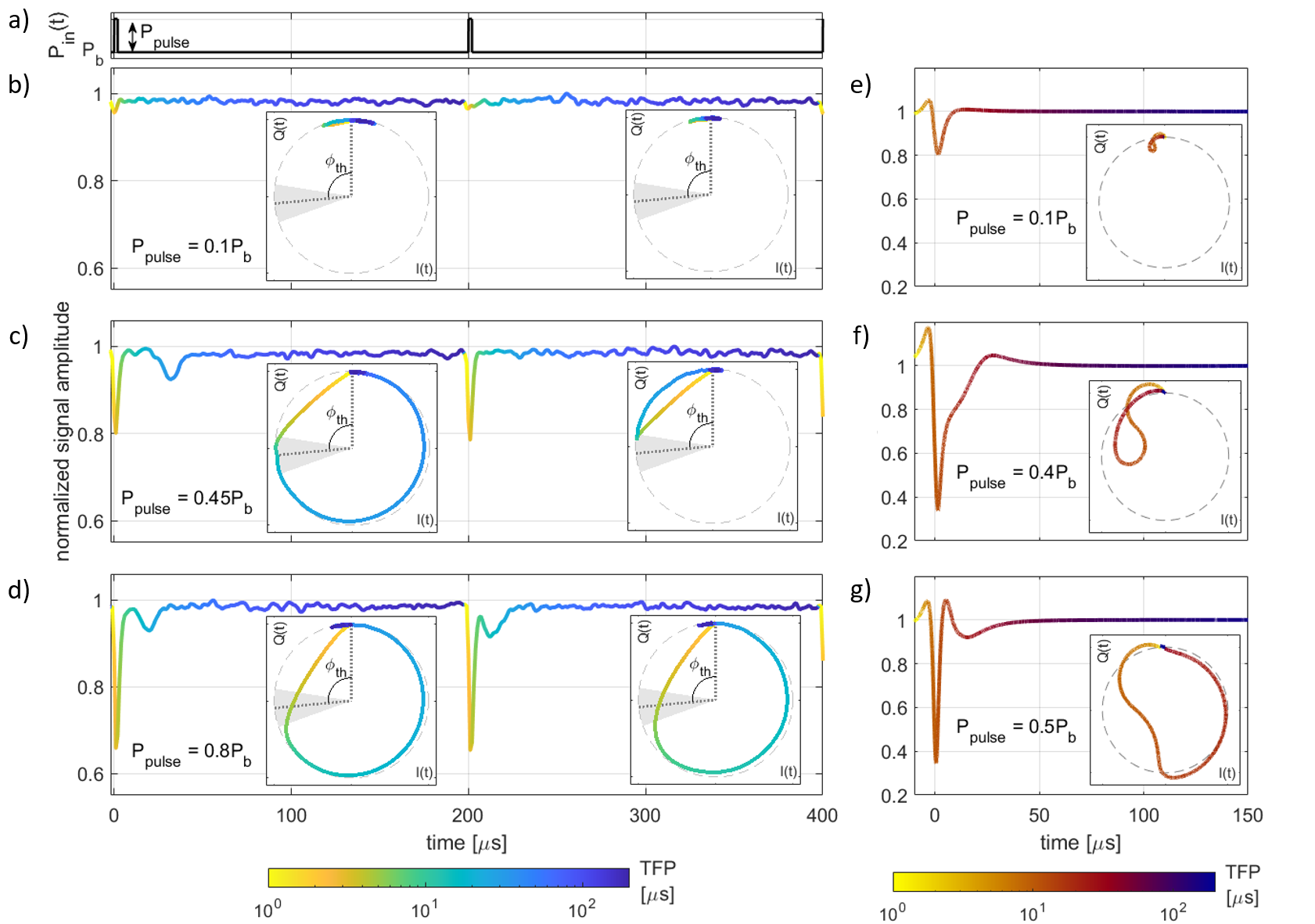}
    \caption{Spiking responses of electro-optomechanical oscillator subject to a perturbation. a) Temporal evolution of the optical input power sent to the device consisting of a bias, cw excitation $P_b = 1.51$ mW, plus a pulse train of rectangular perturbations with $1 \ \mu$s wide pulses of amplitude $P_\mathrm{pulse}$ and $200\ \mu$s period.  b-d) Normalized time evolution of the output signal amplitude recorded by the digital lock-in amplifier after demodulation for $P_\mathrm{pulse}$ equals to (b) $0.1 \times$, (c) $0.45 \times$ and (d) $0.8 \times P_{b}$. The time from perturbation (TFP) is color-coded with a logarithmic scale from the perturbation pulse arrival (in yellow) to 200 $\mu s$ after the perturbation (in dark blue). e-g) Mumerical simulations of the normalized signal amplitude response after numerical demodulation for $P_\mathrm{pulse}$ equals to (e) $0.1 \times$, (f) $0.4 \times$ and (g) $0.5 \times P_{b}$ with the time evolution color-coded in the same way as in the experiment. In insets: phase space reconstructed from the demodulated quadratures $I(t)$ and $Q(t)$. The gray dotted circle represents the phase space dynamics of a pure phase oscillator in the unlocked regime. The dashed black lines in the experimental phase space diagram  mark the average estimated positions of the stable and unstable phase equilibria and their difference $\phi_\mathrm{th}$, for the specific experimental conditions. 
    }
    \label{fig:Fig2_TimeTraces}
\end{figure}

  In order to study the excitability of the E-OMO, the RF detuning is set inside the locking region in the vicinity of its low-frequency edge ($f_{d} - f_{-} = 6$ kHz) while the wavelength of the optical input is kept at $\lambda_o = 1550.5$ nm. Optical perturbations are imprinted onto the input optical field thanks to a Lithium-Niobate electro-optic modulator (EOM in figure \ref{fig:Fig1_SetUp_Lockin}b) controlled by an arbitrary function generator (AFG in figure \ref{fig:Fig1_SetUp_Lockin}b) and a DC electric bias. 
The optical perturbations ($P_{pulse}$) are  set as a pulse train of rectangular pulses having 1 µs duration with a duty cycle of 0.5 $\%$, sitting on-top of a DC optical bias ($P_b$) such that $P_{in}(t)=P_b+P_\mathrm{pulse}$. The time trace of the perturbations is illustrated in figure \ref{fig:Fig2_TimeTraces}a for two periods. The optical response of the system is displayed in  Figure \ref{fig:Fig2_TimeTraces}b--d for three different values of the perturbation pulse $P_\mathrm{pulse}$, namely 10$\%$, 45$\%$ and 80$\%$ of $P_{b}$ respectively.
  The time elapsed since the perturbation is color-coded with a logarithmic scale from 0 (yellow) to 200 µs (dark blue).
  Moreover, the corresponding phase portrait of the transmitted optical signal (with quadratures $I(t)$ and $Q(t)$) are shown as insets where the colors are consistent with the time traces. The gray dotted circle is a guide for the eyes, highlighting the phase space that the system would explore if unlocked to the external RF driving and with no amplitude dynamics (pure phase oscillator case).
   
The mechanism leading to phase perturbations can be understood by looking at figure \ref{fig:Fig1_SetUp_Lockin}c. An increase of the input intensity $P_{in}$ induces a shift of the oscillation frequency. Thus, the optical power perturbations cause abrupt and intermittent phase kicks to the system. As the amplitude of the perturbation increases, the phase shift gets larger till a point where it can exceeds a phase threshold $\phi_\mathrm{th}$ (see Figs. \ref{fig:Fig2_TimeTraces}b-d).
In the case of a weak perturbation ($ 0.1 \times P_{b}$, Fig. \ref{fig:Fig2_TimeTraces}b), practically no amplitude response is visible and the E-OMO remains locked as can be seen in the phase space representation, where the system stays close to the stable locked phase. Small deviations from the stable locked phase are nevertheless evidenced because  of environmental and experimental noises. Note also that an upward perturbation pulse in the input optical power translates here into a dip in the optical response envelope dynamics.
For a slightly larger amplitude perturbation strength ($0.45 \times P_{b}$, Fig. \ref{fig:Fig2_TimeTraces}c), close to the locked-unlocked perturbation threshold $\phi_\mathrm{th}$, the first perturbation induces a response a few tenths of microseconds later, evidenced by the presence of a downward peak in the signal amplitude, unlike the second one. 
From the inspection of the phase space, we observe that the first perturbation is strong enough to unlock the system and induce a $2\pi$, counter clock-wise phase excursion so that the system returns to its original stable locked state (left inset). By contrast, the second perturbation does not produce a full $2\pi$ phase excursion (right inset). Rather, the system settles back to the stable locked phase in the clockwise direction without any noticeable amplitude response.
For an even stronger perturbation amplitude ($0.8 \times P_{b}$, Fig. \ref{fig:Fig2_TimeTraces}d), each perturbation is sufficient to excite a response pulse as evidenced by the presence of downward peaks following both perturbations. In the two phase space representations (see insets), it can be noticed that the perturbations are strong enough to induce a $2\pi$ phase excursion before the system settles back to its original locked state.
This behavior is very typical of an excitable response in a SNIC scenario\cite{Izhikevich_2006}. There exists a clear threshold in the optical perturbation amplitude which produces a full $2\pi$ phase excursion in the response. Besides the $2\pi$ phase excursion, the response is associated with an amplitude response (a small downward pulse) since our E-OMO is not a pure phase oscillator but possesses an amplitude dynamics too. Close to the excitable threshold (Fig. \ref{fig:Fig2_TimeTraces}c), noise can trigger or not a response. However, the response is always of the all-or-none type, since the phase excursion is either a full $2\pi$ excursion  or a return back to the original locked phase, and the amplitude response is either a calibrated downward pulse or practically none. For perturbations well above the threshold (Fig. \ref{fig:Fig2_TimeTraces}d), the response is always present and remains calibrated in phase or in amplitude. In the phase space representation, the transition from clockwise return to locked phase and full $2\pi$ counter-clockwise phase excursion reveals the presence of an unstable fixed point, which is marked by a dashed line in the experimental phase-space representations associated to a gray segment that represents the calculated experimental uncertainty in its position, obtained fron the time evolutions of the recorded phase (not shown here). In particular this allows to reconstruct the relative phase difference between the stable and unstable fixed points, which are reppresented by the angle $\phi_{th}$ in every experimental insets.

Numerical simulations of the optically perturbed E-OMO are carried out with parameters extracted from the experiment (see Methods for details on the modelling and numerical simulations) and their results are displayed in Figure \ref{fig:Fig2_TimeTraces} e--g. The plotted signal is numerically demodulated in the same way as in the experiment to allow for direct comparison. 
Figure \ref{fig:Fig2_TimeTraces} e--g shows the time evolution of the normalized signal amplitude for three distinct perturbation amplitudes $P_\mathrm{pulse}$. For a small perturbation amplitude ($P_\mathrm{pulse}=0.1 \times P_b$ - Fig. \ref{fig:Fig2_TimeTraces}e), which appears similarly as in the experiment as a downward signal, the E-OMO is not perturbed  strong enough to induce a full $2\pi$ phase excursion. Rather, a small dynamics near the stable locked phase is visible in phase-space, as a consequence of the small input perturbation. Note that the simulations are performed without added noise, and are thus purely deterministic. For an intermediate perturbation amplitude ($P_\mathrm{pulse}=0.4\times P_b$ - Fig. \ref{fig:Fig2_TimeTraces}f), close but below the excitable threshold, an asymmetric response is visible without a complete excursion in phase space. For an even larger perturbation ($P_\mathrm{pulse}=0.5\times P_b$ - Fig. \ref{fig:Fig2_TimeTraces}g), two distinct downward pulses separated by about $15 \mu s$ appear. In the phase space representation and with the color-coding of the time from perturbation, it appears that the second pulse is associated with the $2\pi$ phase excursion, evidencing the numerical excitable response.
The good qualitative agreement with the experimental recordings validates the model used here, and confirms the excitable response experimentally observed. The only difference here is that no noise has been injected in the numerical simulations, therefore no noise excited excitable pulse can be observed. 

\section{Characterization of excitable response main properties}

\begin{figure}
    \centering
    \includegraphics[width=1\linewidth]{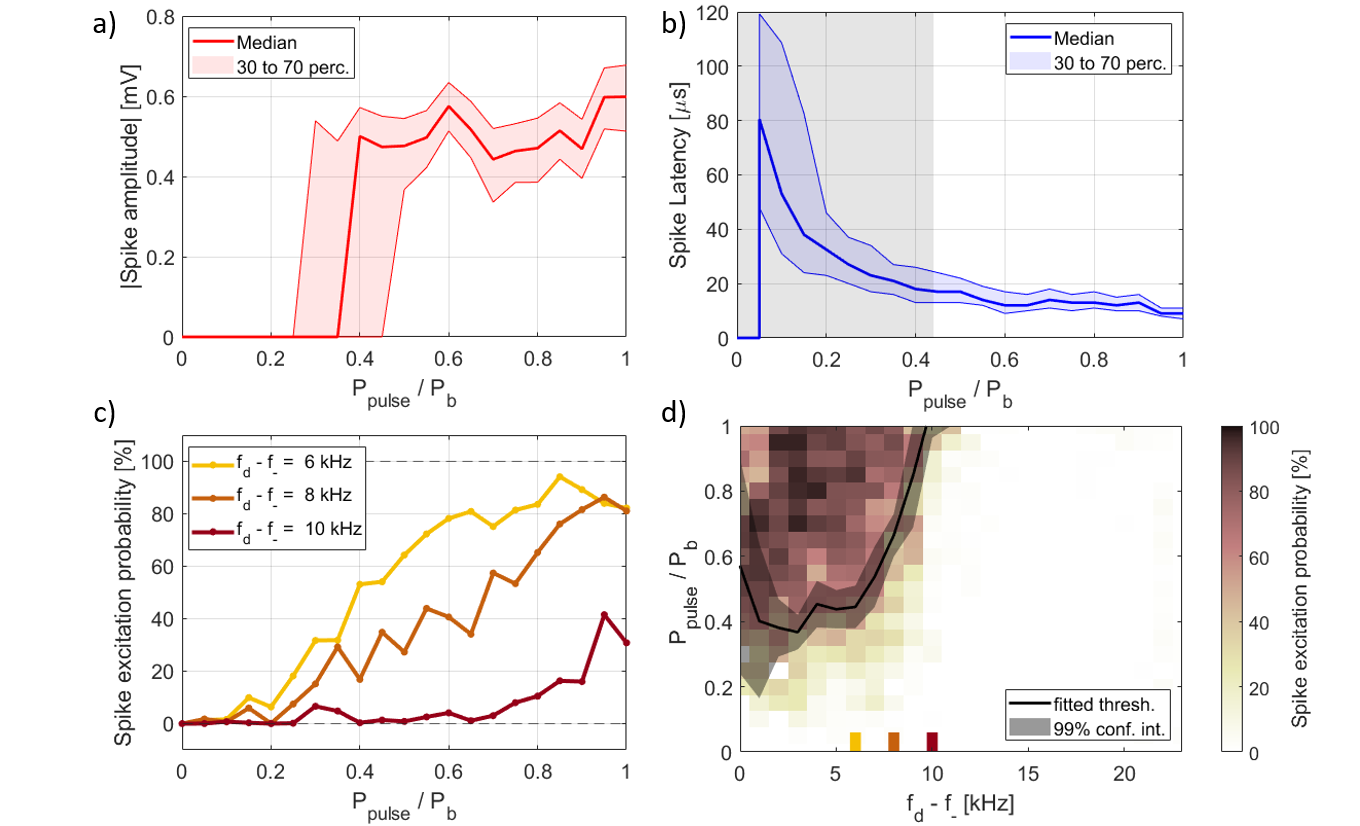}
    \caption{Statistical analysis of the excitable response as a function of the perturbation amplitude. a) Median amplitude of the spiking response in absolute value, at $f_{d} - f_{-}=6$ kHz. The shaded region refers to the 30$^{th}$ and 70$^{th}$ percentiles. b) Median spike latency, at $f_{d} - f_{-}=6$ kHz. The shaded blue region refers to the 30$^{th}$ and 70$^{th}$ percentiles whereas the gray shaded one refers to the below-threshold perturbation amplitudes, $P_{pulse}\gtrsim0.44\times P_{b}$. c) Spike excitation probability as a function of the perturbation  amplitude for three different values of  $f_{d} - f_{-}$ : 6 kHz (yellow), 8 kHz (orange) and 10 kHz (dark red). d) Map of the spike excitation probability (color-scale) as a function of the perturbation amplitude and of the RF drive detuning  $f_{d} - f_{-}$.
    }
    \label{fig:Fig3_Excitability}
\end{figure}

The excitable threshold as well as the time delay between the perturbation and the excitable response (spike latency) are characterized by perturbing 5000 times (overall time traces of 1 second) the locked state with the same parameters ($P_{b} = 1.51$ mW and $f_{d} - f_{-} = 6$ kHz) and performing a statistical analysis of the response.
Fig. \ref{fig:Fig3_Excitability}a shows the median of the absolute value of the response amplitude to a single perturbation with 30-70 percentiles. For low perturbation powers until $P_\mathrm{pulse}\simeq 0.35 \times P_{b}$, the median of the response amplitude is 0 since the perturbation never or rarely gives rise to an excitable response. Above this threshold ($P_\mathrm{pulse}\gtrsim 0.4\times P_{b}$), excitable pulses appear with a calibrated amplitude response median, a sign of the calibrated excitable response pulse whose amplitude hardly depends on the perturbation strength. This is a clear signature of the expected all-or-none response, with a marked threshold.
Fig. \ref{fig:Fig3_Excitability}b displays the spike latency as a function of the perturbation amplitude $P_\mathrm{pulse}$. It shows a highly nonlinear evolution, with low latencies associated to high perturbation amplitudes followed by a rapidly increasing latency around the excitable threshold till a zero latency corresponding to no emitted pulse at all. It could be found at least curious or even wrong that latency times could be extracted below the excitable threshold, in the grayed region in fig. \ref{fig:Fig3_Excitability}b). However, this is easily understood when considering the noise in the system. Indeed, for non-zero perturbation amplitude $P_\mathrm{pulse}$ but below the excitable threshold (in the gray area), a median for the spike latency can still be extracted because of frequent noise-induced pulses, until a point where the noise is no more strong enough to induce noise-excited excitable responses. The overall latency is well inline with what is expected in the excitable regime, with a divergence of the latency time at the excitable threshold in the deterministic case. A reduction in latency by a factor $4$ is evidenced when the perturbation amplitude increases by a factor about 20, where the spike latency  reaches saturation at 15 $\mu$s above $0.75 \times P_{in}$. This saturation value is of the same order as the one reached in the numerical simulations, as can be observed in Fig. \ref{fig:Fig2_TimeTraces}g. Note that spike latency introduces an important brain-inspired computing mechanism for encoding input perturbations in the time domain of the output spikes \cite{GAUTRAIS199857,StocklNMI21}.

The probability of triggering an excitable pulse is investigated as a function of the optical perturbation amplitude for different RF drive detunings in Fig. \ref{fig:Fig3_Excitability}c. For $f_{d} - f_{-} = 6$ kHz, the probability reaches over 80 $\%$ beyond a perturbation amplitude of $0.75 \times P_{b}$ (Fig. \ref{fig:Fig3_Excitability}c - yellow curve). By going deeper in the locking region, stronger perturbations are needed to reach efficient excitability (above $100\%$ $P_{b}$ for $f_{d} - f_{-} = 8$ kHz - orange curve in fig. \ref{fig:Fig3_Excitability}c), up to a frequency detuning of 10 kHz (red curve in fig. \ref{fig:Fig3_Excitability}c) where no more than 40 $\%$ efficiency can be reached within the linearity range of the EOM.
The probability for an excitable pulse response is fully mapped in the plane of the detuning of the RF drive versus the amplitude of the optical perturbation $P_\mathrm{pulse}$ (figure \ref{fig:Fig1_SetUp_Lockin}d). For each detuning, the evolution as a function of the power ratio is fitted with a sigmoid function. The 50 $\%$ limit is highlighted in black, and the 99 $\%$ confidence interval is highlighted in shaded gray. 
Close to the locking-unlocking transition, below $2$ kHz, a reduced switching efficiency is observed due to an averaging effect of the noise-induced fluctuations of the mechanical self-oscillation frequency. At $2$ kHz detuning, the 50 $\%$ limit is reached around $P_{pulse}=0.4 \times P_{b}$ and the probability of over 90 $\%$ is achieved with a pulse power of $0.7 \times P_{b}$. By further increasing the detuning, the 50 $\%$ limit is slightly reduced till $6$ kHz. Beyond this detuning, the strength of the perturbation amplitude required to reach 50 $\%$  rapidly increases and the excitable efficiency drops. Beyond $10$ kHz, the observation of excitability is no longer possible with the current experimental apparatus and available perturbation amplitudes. However, we expect a strong increase of the perturbation amplitude in this regime and even the disappearance of the excitable response. 
This shift to higher perturbation amplitudes as the detuning of the RF drive gets larger demonstrates the possibility to easily tune and control the excitable threshold. Indeed, keeping $P_{b}$ fixed while increasing the detuning ($f_{d} - f_{-}$) increases the excitable threshold, making excitable pulses harder to excite. This simple behaviour is important in the context of analog brain inspired computing, as e.g. demonstrated in a photonic context in \cite{Masominia_2023}.

\section{Summation and Refractory periods}

\begin{figure}
    \centering
    \includegraphics[width=1\linewidth]{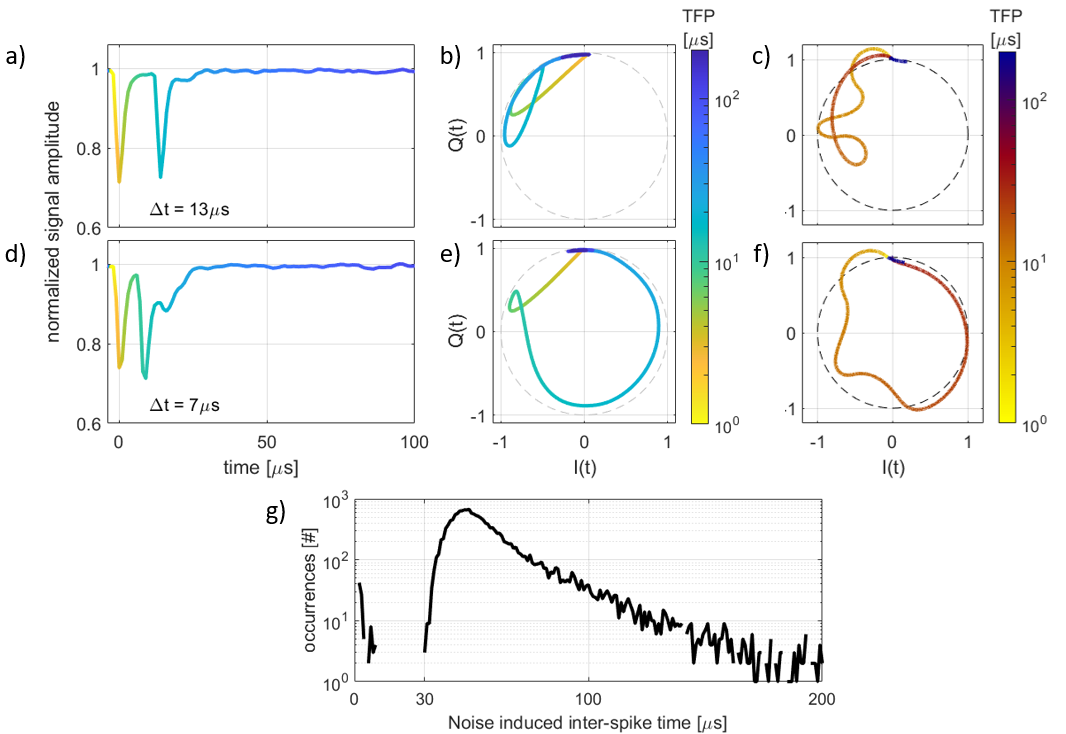}
    \caption{Temporal summation and refractory period dynamics. a,d) Experimental time traces and b,e) phase spaces in the cases of two sub-threshold perturbations separated by a time delay $\Delta t=13 \ \mu$ s and $\Delta t=13 \ \mu$ s respectively. c,f) Numerical simulations of the dynamics in the phase space with two sub-threshold perturbations arriving with time delays $\Delta t=6 \ \mu$ s and  $\Delta t=4 \ \mu s$ respectively.  g) Experimental distribution of the noise excited inter-spikes time for $f_d-f_-=0$ kHz. }
    \label{fig:Fig4}
\end{figure}

Beyond the excitable response, biological neurons show other important features such as temporal summation and refractory periods. The ability of neurons to integrate sub-threshold input signals received on their input channels can be probed by studying the dynamical response to consecutive below-threshold perturbations depending on their separation in time \cite{Selmi_2015}. 
Two equal and below-threshold optical perturbation pulses ($P_{pulse}=0.4\times P_{b}$ and $1 \ \mu s$ pulse width) with a controllable time delay $\Delta t$ are prepared. For a time delay $\Delta t=13 \ \mu s$, figure \ref{fig:Fig4}a shows the normalized signal amplitude in time while the corresponding phase space is displayed in Fig. \ref{fig:Fig4}b. It is clear from the two plots that the two consecutive perturbations do not induce an excitable response.
However, by reducing the time delay down to $\Delta t=7 \ \mu s$, the two perturbations can add up so that an excitable response can be observed in the normalized signal amplitude (fig. \ref{fig:Fig4}d) after the second perturbation. The corresponding phase space (fig. \ref{fig:Fig4}e) highlights the $2\pi$ phase shift evidencing the excitable behavior for two sub-threshold perturbations.
This behaviour is well reproduced in the numerical simulations of our model in \ref{fig:Fig2_TimeTraces}e--g, where two successive square wave optical power perturbations are introduced as inputs into the cavity. For below threshold perturbations ($P_{pulse}=0.3 \times P_{b}$ and $f_{d}-f_{-}=15 \ kHz$), simulated phase spaces for $\Delta t=6 \ \mu s$ and $4 \ \mu s$ are shown in figure \ref{fig:Fig4}c and f respectively. In the first case (Fig. \ref{fig:Fig4}c) the pulses are too far away apart to trigger an excitable response and a $2\pi$ phase excursion. In the second case, (Fig. \ref{fig:Fig4}d), a $2\pi$ phase excursion is produced.
 %This is due to the low pass filtering effect combined with the presence of many features close to each other (the $2$ perturbations and possibly the spike).

Another important feature of neurons concerns the response to successive, supra-threshold perturbations. Once an supra-threshold perturbation excites a spike, the neuron is not able to respond to another perturbation before a given period of time called refractory time or period \cite{Purves_2001}.
%This effect describes two characteristics of the neuron, it limits the number of action potentials that can be produced per unit time and it explains why the action potentials do not propagate backward toward the point of their initiation as they travel along an axon. \\
One method to extract the refractory period is to investigate the inter-spike time distribution induced by external noise sources \cite{Eguia_2000, Pedaci_2011}.
Such distribution is shown in Fig. \ref{fig:Fig4}g for $f_d-f_{-} = 0$ kHz. As expected, when noise-induced transitions occur, the inter-spike time distributions follow an exponential decay given by a Kramer's type escape rate \cite{KramersP40}. Interestingly, at short time scales, the inter-spike time distribution shows a pronounced dip, due to the impossibility to excite two response spikes that are arbitrarily  close in time. This allows to extract the refractory period of the system  which is of the order of $30\ \mu s$.
%Note that the few spike occurrences calculated at an interspike time $\sim0\ \mu s$ are due to numerical errors while evaluating if the spike has been excited or not.\\

% \begin{figure}
%     \centering
%     \includegraphics[width=1\linewidth]{Images//FIG_4InterSpikeTimes.png}
%     \caption{Evolution of the number of noise-induced occurences as function of the inter-spikes time for values of $f_d-f_{-}$ equals to 0 $kHz$ (a - yellow), 1 $kHz$ (b - orange), 2 $kHz$ (c - dark orange), 3 $kHz$ (d - brown).}
%     \label{fig:FIG_4InterSpikeTimes}
% \end{figure}

\section{Conclusion}
%We have demonstrated an optomechanical spiking neuron with Si integrated optical and electromechanical interfaces, operating at 1550 nm for the photonic interface and in the 3 GHz range for the mechanical one, paving the way for a novel nano-optomechanical neuromorphic platform.
%This platform may open great opportunities for all Si integrated spike-based processing of optical or mechanical signals, but also for edge processing applications, where the data processing takes place near their source rather than remotely, thus limiting bandwidth requirements and latency. It could also have impact in energy efficient generation of arbitrary, on-chip optical spike sequences  instead of current optical modulators based on Mach-Zehnder structures and necessitating high driving voltages, and thanks to the small footprint and high integration of our device.
%In addtion, optical and RF input ports in the electro-optomechanical neuron allow for various injection schemes  yet to be explored. Some of them may open new avenues for the demonstration of on-chip cascadability between similar RE-OMO artificial neurons. Moreover, the integration on photonic integrated circuit could also allow the optical manipulation the phase of the mechanical oscillator {\bf [je ne comprends pas ici]}. Finally, the different optical and mechanical modes naturally present in our nanobeam structure may open opportunities for optical or mechanical multiplexing for future analogue computing.

We have demonstrated an optomechanical spiking neuron based on silicon-integrated optical and electromechanical interfaces. The device operates at telecommunication wavelengths (1550 nm) and exploits a mechanical mode in the 3 GHz range, establishing a new nano-optomechanical platform for neuromorphic information processing.

This platform enables fully silicon-integrated, spike-based processing of optical and mechanical signals and is particularly well suited for edge-computing applications, where data processing is performed close to the signal source, thereby reducing latency and bandwidth demands. Beyond signal processing, the device can provide an energy-efficient route to the generation of arbitrary on-chip optical spike sequences, offering a compact alternative to conventional Mach–Zehnder-based optical modulators, which typically require large footprints and high driving voltages.

The coexistence of optical and radio-frequency input ports enables a broad range of injection and control schemes that remain to be explored. These capabilities may enable on-chip cascadability between multiple electro-optomechanical neurons, a key requirement for scalable neuromorphic architectures. Finally, the multiple optical and mechanical modes inherently supported by the nanobeam structure open opportunities for optical and mechanical multiplexing and parallel neuromorphic computing systems.

%\textcolor{red}{comment on the speed ?}

%{\color{red}[I would like to comment on the scalability and cascadability but we would need to demonstrate at least in the numerics. At the moment, one drawback of the optical injection is that the injected pulse is partially transmitted. Thus the optical signal is "polluted". I don't see why we could not perturb first on the electronics and try to cascade the optical excitable pulse to another system to elicit an excitable response. This would be a start. Another interesting prospect that can be checked in the numerics is the manipulation of the phase of the mechanical vibration. Most probably, the phase excursion seen in the optical signal is present in the mechanical signal. It means we can optically manipulation the phase of the mechanical oscillator !]}

%{\color{red}[I would remove this: Besides the potential interest of directly interfacing this system with mechanical signals in the GHz frequency range as input information to be processed, the developped optomechanical neuron will have the advantages of (i) being fast with the ability to reach few tens of nanoseconds timescale spike duration, (ii) possessing an optical interface (addressing and readout) and (iii) being integrated on chip, scalable, and energy efficient.]} {\color{red}[All this I find dangerous because we don't demonstrate ns spikes, neither scalability or energy efficiency. ]}

\bmhead{Acknowledgements}

This work was funded by ANR project OptoMecaSNIC (ANR-24-CE24-3452) and PEPR Électronique - OROR (ANR-22-PEEL-0008). This work was
partially supported by the French Renatech network.

\section{Methods}
\paragraph{Sample fabrication} 
A 300 nm thick gallium phosphide (GaP) layer with a 280 nm spacer layer of Silicon Nitride (SiN) is heterogeneously integrated on top of Silicon-on-Insulator (SoI) circuitry by wafer bonding. The optomechanical crystal as well as the surrounding pads are first patterned on GaP by electronic beam lithography and dry-etching techniques. The 1 µm wide metallic electrodes are subsequently drawn on both sides of the crystal to ensure electromechanical excitation. Their relative position is chosen to be 1.5 µm away from the optical cavity to avoid degradation of its optical quality factor. Finally, dry-etching of the underlying SiN layer is performed allowing simultaneously a release of the optomechanical crystal and a fine control of the evanescent coupling between the SoI waveguide and the optical cavity.

\paragraph{Data acquisitions}
The optical signal at the output of the integrated device is transduced into an electrical one by a $3.5$ GHz bandwidth photo-detector. This signal is then mixed with the internal reference frequency $f_d$, and low-pass filtered with a demodulation bandwidth of $100$ kHz. The data sampling rate of $1$ MSa/s allows for the recording of a $1$ s long time trace with a time resolution of $1$ µs inside which is recorded the E-OMO excitable response to $5000$ equal input power perturbations.
% The experimental amplitude and phase space dynamics, in fig. \ref{fig:Fig2_TimeTraces}b-d and fig. \ref{fig:Fig4}a,b,d,e, are recorded by the digital lock-in amplifier.\\ 
% by the lock in amplifier  The plotted normalized amplitude and phase spaces derive from the mixing and low pass filtering of the optical signal at the output  with a demodulation bandwidth of $100$ kHz and data sampling rate of $1$ MSa/s allowing for a time resolution of $1\ \mu s$

\paragraph{Numerical model}
The theoretical model consists of three coupled ODEs for the intracavity optical field $a(t)$, the mechanical mode displacement $x(t)$ and the nanobeam temperature elevation $\Delta T(t)$ and reads:

\begin{eqnarray*}
\dot{a}(t) & = & \left[i\left(\Delta\omega+g_{0}x(t)+\frac{\omega_{L}}{n_{0}}\kappa_{th}\Delta T\right)-\frac{\kappa}{2}\right]a(t)+\sqrt{\kappa_{ex}}s_{in}(t)\label{eq:a(t)_with_g0}\\
\ddot{x}(t) & = & -\Gamma_m\dot{x}(t)-\Omega_{m}^{2}x(t)+2\Omega_{m}g_{0}|a(t)|^{2}+F_{0}sin(2\pi f_{D}t)\label{eq:x_norm_Frp}\\
\dot{\Delta T(t)} & = & \frac{\Gamma_{th}}{\tau_{th}}|a(t)|^{2}-\frac{\Delta T(t)}{\tau_{th}}\label{eq:DT_OMOeq}
\end{eqnarray*}

where $s_{in}(t) = \sqrt{\frac{P_{in}}{\hbar\omega_L} \left( 1+P_{pulse}S(2\pi f_Pt,DC) \right)}$ represents the input optical field. $P_{in}=0.38$ mW is the input power in the bus waveguide to excite the OM self sustained oscillations, $P_{pert}$ the perturbation pulse height with respect to $P_{in}$ and $S(2\pi f_Dt,DC)$ is the square wave signal with amplitude $\in[0, \ 1]$, frequency $f_P=5 \ kHz$ and duty-cycle DC.
The optical mode amplitude $a(t)$ is driven by the input field $s_{in}(t)$ with a coupling constant $\kappa_ex$. The detuning between the input field and the cavity resonance is $\Delta \omega = \omega_L-\omega_{cav}$. The optomechanical coupling is $g_0x(t)$ and the optical cavity is also subject to a thermally induced detuning where $n_0$  is the cavity material refractive index and $\kappa_{th}$ the thermo-optic coefficient.  The field decay rate $\kappa$ is calculated as $\kappa=\omega_{cav}/Q_o$ where $Q_o$ is the optical resonance loaded quality factor. The mechanical displacement $\tilde{x}$ is normalized to the the zero-point-fluctuation displacement  such that $x(t)=\tilde{x}(t)/x_{zpf}$, where $x_{zpf}=\sqrt{\hbar/2m_{eff}\Omega_m}$, with $m_{eff}$  the mechanical oscillator effective mass and $\Omega_m$ its natural resonant frequency. 
The transmitted electric field intensity is calculated as $S_T(t)=|s_T(t)|^2=|s_{in}(t)-\sqrt{\kappa_{ex}}a(t)|^2$.\\

The mechanical mode is described by the equation for a damped harmonic mechanical oscillator, that is driven by radiation pressure force and the piezoelectrical actuation $F_0=F/m_{eff}x_{zpf}$,  with $F$ the driving force applied on the mechanical oscillator and $f_D$ the driving frequency.
$\Gamma_m = \Omega_m/Q_m$ where $Q_m$ is the mechanical quality factor.
The temperature increase of the cavity region $\Delta T(t)$  can be described by a source term which depends on the cavity absorption,  and decays  with a characteristic time $\tau_{th}$, following \cite{guha_high_2017}.  
The thermal load $\Gamma_{th} = R_{th}\hbar\omega_L\kappa_{abs}$ depends on the thermal resistance $R_{th}=\Delta x / k_{GaP}A$  with $\Delta x$ the path parallel to the heat flow, $k_{GaP}$ the thermal conductivity of gallium phosphide and $A$ the cross section perpendicular to the heat flow. $\kappa_{abs}$ is the decay rate due to photon absorptions in the cavity region, estimated as $\kappa_{abs} = 0.1\kappa_{in}$ \cite{zhang_multibistability_2013}.

The full parameters used for the numerical simulations are given in the Table \ref{tab:values_OMexcit} below. 

\begin{table}
    \centering
    \begin{tabular}{|c|c|c|c|c|} \hline 
 $Parameter$& $Value$ &$Units$&$Comments$ &$Ref.$\\ \hline 
         $\lambda_{cav}$&  $1549$&$nm$& cavity resonant wavelength &measure\\ \hline 
         $g_0/2\pi$&  $660$&$kHz$& optomechanical coupling rate&measure\\ \hline 
         $n_0$&  $3.05$ &-& GaP refractive index @ $\lambda = 1550 \ nm$ &\cite{tian_piezoelectric_2024}\\ \hline 
         $\kappa_{\Delta T}$&  $3.4 \times 10^{-5} \ $&$K^{-1}$& GaP TO coefficient &\cite{lake_efficient_2016}\\ \hline 
         $Q_o$&  $40000$&-& optical Quality factor &measure\\\hline
 $\kappa$& $2\pi c/\lambda_{cav}Q_O$& $s^{-1}$& cavity field decay rate&calc.\\\hline
 $\kappa_{ex}$& $\kappa/3$& $s^{-1}$& external decay rate&free param.\\ \hline 
         $Q_m$&  $1000$ &-& mechanical quality factor &measure\\ \hline 
 $\Omega_m/2\pi$& $3078$&$MHz$&mechanical frequency &measure\\ \hline 
 $F$& $50$&$pN$&external forcing&free param.\\ \hline 
 $m_{eff}$& $10^{-16} \ $&$kg$&effective mass &measure\\ \hline 
 $\Delta x$& $40 \ $&$\mu m$&cavity length&design\\ \hline 
 $k_{GaP}$& $110 \ $&$W/mK$&GaP thermal conductivity &\cite{levinshtein_handbook_1999}\\ \hline 
 $A$& $300 \times 700 $&$nm^2$&cavity cross section&design\\ \hline
 $\tau_{th}$& $0.3$&$\mu s$&thermal decay time const. &\cite{chan_optimized_2012, navarro-urrios_self-stabilized_2015}\\ \hline
    \end{tabular}
    \caption{Table of values for electro-optomechanical equations}
    \label{tab:values_OMexcit}
\end{table}

\break
\bibliography{OM_Excitability_BIB}

\end{document}